\documentclass[letterpaper, 10 pt, conference]{ieeeconf}  

\IEEEoverridecommandlockouts                              

\usepackage{graphics} 
\usepackage{epsfig} 
\usepackage{mathptmx} 
 \usepackage{amsmath,bm}
 
 \usepackage{amsfonts}
\usepackage{amssymb}
\usepackage{times} 
 \usepackage{graphicx}
\usepackage{xcolor}

\newtheorem{theorem}{Theorem}
\newtheorem{asum}{A}
\newtheorem{rem}{Remark}
\newtheorem{prop}{Proposition}

\title{\LARGE \bf
Data-Driven Feedback Linearization of Nonlinear Systems with Periodic Orbits in the Zero-Dynamics
}

\author{Karthik Shenoy$^{1}$, Akshit Saradagi$^{2}$, Ramkrishna Pasumarthy$^{1}$, Vijaysekhar Chellaboina$^{3}$
\thanks{$^{1}$KS  and RP  are with the Dept. of Electrical Engineering, IIT-Madras, India.
        {\tt\small \{ee21d405@smail, ramkrishna@ee}\}.iitm.ac.in}
\thanks{$^{2}$AS is with the Department of Computer Science, Electrical and Space Engineering, Lule\aa~University of Technology, Sweden. 
        {\tt\small akssar@ltu.se}}
\thanks{$^{3}$ VC is the Dean of Engineering, GITAM (Deemed to be University), Visakhapatnam, India. 
        {\tt\small dean\_engineering@gitam.edu}}%
}

\begin{document}

\maketitle
\thispagestyle{empty}
\pagestyle{empty}

\begin{abstract}
 In this article, we present data-driven feedback linearization for nonlinear systems with periodic orbits in the zero-dynamics. This scenario is challenging for data-driven control design because the higher order terms of the internal dynamics in the discretization appear as disturbance inputs to the controllable subsystem of the normal form. Our design consists of two parts: a data-driven feedback linearization based controller and a two-part estimator that can reconstruct the unknown nonlinear terms in the normal form of a nonlinear system. We investigate the effects of coupling between the subsystems in the normal form of the closed-loop nonlinear system and conclude that the presence of such coupling prevents asymptotic convergence of the controllable states. We also show that the estimation error in the controllable states scales linearly with the sampling time. Finally, we present a simulation based validation of the proposed data-driven feedback linearization.   
\end{abstract}

\section{INTRODUCTION}
Most classical control techniques rely on the availability
of a plant model to design controllers. The
models can either be derived through first principles or
via system identification techniques. As systems get larger
and processes become more complex deriving and working
with first principle models may no longer be feasible. 
 With the advent of advanced sensing technologies, these complex processes generate and store large amounts of data, which can be used to design controllers \cite{brunton}, effective decision-making \cite{Dec}, and predicting and diagnosing faults \cite{fault}.   
A plethora of literature is available for data-driven control of linear systems. 
Based on the Willems' fundamental lemma, \cite{formulabased} shows that in the presence of persistently exciting inputs, a state feedback control law can be derived without explicitly identifying the system model. 
Data-driven techniques have since been extended to deriving control laws such as event-triggered control \cite{ETC}, LQR \cite{LQR}, minimum-energy control \cite{minen}, passivity-based techniques using measured data \cite{dissipativity} and time delay systems \cite{delay} among several others. The literature for data-driven methods for nonlinear systems is sparse in comparison.  In \cite{NLdatadriven}, the authors lift an unknown nonlinear system dynamics to a polynomial space in which the dynamics has a linear structure. The lifted dynamics is then used to design polynomial control laws using data-driven techniques. In \cite{tabuada}, results are derived for data-driven full-state feedback linearization of nonlinear systems, assuming that the systems are available in the normal form. 
The results also extend to the case of partial feedback linearization when the zero dynamics is asymptotically stable. An exciting case of data-driven partial feedback linearization is when the zero dynamics is not asymptotically stable but has a stable periodic orbit. Such scenarios are encountered in systems like biped robots \cite{walking} and a network of coupled oscillators that exhibit periodic behavior in their zero dynamics. The gait in a biped robot is made to converge to a stable limit cycle during walking. This leads to a periodic behavior in the entire structure, above the hip joint. Thus, stabilizing the head and the torso of the biped robot, using feedback linearization, can be categorized as an input-output linearization problem with a periodic behavior in the zero-dynamics. Similar cases can be found in cluster synchronization \cite{Cluster} of a network of coupled nonlinear oscillators using feedback linearization, such that certain sub-clusters in the zero dynamics converge to a stable periodic orbit.

A limit cycle in the zero dynamics behaves like a periodic disturbance that is coupled to the \emph{controllable} subsystem (we discuss how the system can be transformed into its normal form, that partitions the system into a controllable and an internal part in Section \ref{prelimsfb}). When the normal form of the system is known, this can be solved by the exact cancellation of the nonlinear terms, that appear in the controllable subsystem, via feedback. The challenge is when we lack knowledge of these nonlinear terms that appear in the normal form. In such cases, to implement data-driven techniques, and due to the inherent discrete nature of data-driven techniques, we require a sampled-data equivalent of the normal form of the nonlinear system. This, as we will see in the later  sections, will give rise to higher-order terms in the controllable subsystem which cannot be canceled via feedback linearization. These higher-order terms act as persistent disturbance inputs to the controllable subsystem and inhibit the states from achieving asymptotic convergence. In this paper, we try to address these problems while using data-driven partial feedback linearization when there exists a stable periodic orbit in the zero dynamics. We encapsulate the main contributions of this paper below:
\begin{enumerate}
\item We design a data-driven partial feedback linearization-based controller to stabilize the controllable subsystem of a nonlinear system, with unknown normal forms and those which contain periodic orbits in zero dynamics. We prove the disturbance-to-state stability of the controllable subsystem, where the higher-order terms in the internal dynamics appear as disturbance inputs to the controllable subsystem.      
\item We exemplify the results derived using simulations and critically analyze how the estimation error varies with the sampling time. We also show the robustness of the controller in the presence of perturbations in the assumed model structure.   
\end{enumerate}
\section{Notations}
$\mathbb{R}^{++},\;\mathbb{R}^{+}$ denote the set of positive and non-negative real numbers respectively. $col\{x_1,x_2,\dots,x_n\}$ represents a column vector with the entries $x_1,x_2\ldots,x_n$. Estimates of a state will be represented by a hat on top. For example, $\hat{\xi}$. $||x||$ represents the 2-norm of a vector $x\in\mathbb{R}^n$. The big-$O$ notation: $f(z,t)=O_z(T)$ implies $\exists\;N,T\in\mathbb{R}^+$ such that, $\forall t\in[0,T]$, $||f(t,z)||\leq NT||z||$. $\lambda_{m}(Q)$ represents the minimum eigenvalue of a matrix $Q\in\mathbb{R}^{n\times n}$. $X^{\dagger}$ represents the pseudo-inverse of a matrix $X$. $\mathbb{O}$ represents a matrix of zeros of appropriate dimension. $||A||_F$ represents the Frobenius norm of a matrix. The Lie-derivative of a function $h(x): \mathbb{R}^n\rightarrow\mathbb{R}$ along a vector field $f(x):\mathbb{R}^n\rightarrow\mathbb{R}^n$ is given by $L_fh(x)=\frac{\partial h(x)}{\partial x}\cdot f(x)$.
\section{Preliminaries and Problem Formulation}
\subsection{Persistency of Excitation \cite{formulabased}}\label{poesect}
A signal $u=col\{u(0),\ldots,u(T-1)\}$ where $u(\cdot)\in\mathbb{R}^n$ is persistently exciting of order $G$ if the Hankel matrix:
\begin{align*}
H^{T-G+1}_G(u(0))=
    \begin{bmatrix}
    u(0) & u(1) & \hdots & u(T-G)\\
    u(1) & u(2) & \hdots & u(T-G+1)\\
    \vdots & \vdots & \ddots & \vdots\\
    u(G-1) & u(G) & \hdots & u(T-1)
    \end{bmatrix}
\end{align*}
has a full rank of $nG$ with $T\geq(n+1)G-1$.
\subsection{Feedback Linearization} \label{prelimsfb}
Feedback linearization \cite{slotine}, \cite{isidori} is a technique by which a nonlinear system can be transformed into a linear system via a proper choice of a nonlinear transformation and a  state feedback control law. Consider the SISO system
\begin{equation}\label{flsys}
    \begin{aligned}
    \dot{z}&=f(z(t))+g(z(t))u(t),\;\;y(t)=h(z(t))
\end{aligned}
\end{equation}
where $f:D\rightarrow\mathbb{R}^n$, $g:D\rightarrow\mathbb{R}^n$ and $h:D\rightarrow\mathbb{R}$ are sufficiently smooth in the domain $D\subset\mathbb{R}^n$, $u,y\in\mathbb{R}$. System \eqref{flsys} is said to have a relative degree $\rho$, $1\leq\rho\leq n$ in a region $D_0\subset D$ if for $i=1,2,...,\rho\mbox{--}1$, $L_gL_f^{i\mbox{--}1}h(z(t))=0$ and $L_gL_f^{\rho\mbox{--}1}h(z(t))\neq0$.
for all $z(t)\in D_0$. Consider the general case, where $\rho\leq n$ at $x^0$. Then, choosing 
\begin{align*}
\xi(t)=
\begin{bmatrix}
 \xi_1(t)\\\xi_2(t)\\\vdots\\\xi_{\rho}(t)
\end{bmatrix}=\begin{bmatrix}
 h(z(t))\\L_fh(z(t))\\\vdots\\L_f^{\rho-1}h(z(t))
\end{bmatrix}=\Phi(z(t))
\end{align*}
together with functions $\psi_{1},\dots,\psi_{n-\rho}$, to transform the rest of the $n-\rho$ states gives us the dynamics in a new set of coordinates $\begin{bmatrix}
     \xi(t)\\\hline\eta(t)
    \end{bmatrix}=\begin{bmatrix}
     \Phi(z(t))\\\hline\Psi(z(t))
    \end{bmatrix}$ transforms system \eqref{flsys} into 
\begin{align} 
    \dot{\eta}(t)&=f_0(\eta(t),\xi(t)) \label{plant1}\\ 
    \dot{\xi}_{i}(t)&=\xi_{i+1}(t),\;\;for\;\;1\leq i\leq \rho-1 \label{plant2}\\ 
    \dot{\xi}_{\rho}(t)&=\alpha(\xi(t),\eta(t))+\beta(\xi(t),\eta(t)) u(t) \label{plant3}\\
    y(t)&=\xi_1(t) \label{plant4}
\end{align}
where 
 $\alpha(\xi(t),\eta(t))=L_gL_f^{\rho-1}h(z(t))$ and $\beta(\xi(t),\eta(t))=\dfrac{L_f^{\rho}h(z(t))}{L_gL_f^{\rho-1}h(z(t))}$. Here the system is partitioned into its \emph{controllable states} $\xi(t)$ and \emph{internal states} $\eta(t)$. Now by choosing a control law of the form $u(t)=\beta^{\mbox{--}1}(\xi(t),\eta(t))(v(\xi(t))-\alpha(\xi(t),\eta(t)))$,
 where $v(\xi(t))$ is any linear control law, we get the linear the input-output dynamics:
\begin{equation}\label{fblinsys}
    \begin{aligned}
         \dot{\eta}(t)&=f_0(\xi(t),\eta(t))\\
    \dot{\xi}(t)&=A_c\xi(t)+B_cv(\xi(t)),\;\;y(t)=C_c\xi(t)
    \end{aligned}
\end{equation}
where $A_c,B_c,C_c$ are the matrices corresponding to the chain of integrators in the Brunovsky’s canonical form.
The zero dynamics of \eqref{fblinsys} is given by $\dot{\eta}(t)=f_0(\eta(t),0)$. The zero dynamics of the system, which is unobservable, dictates whether \eqref{plant1}-\eqref{plant4} can be stabilized by the control input $u(t)$. Such systems, whose zero dynamics is asymptotically stable are called minimum-phase systems. Henceforth, the time argument $t$ in the states will be omitted and will be explicitly mentioned only when required.
\subsection{Problem Formulation and Assumptions}
Firstly, the classical feedback linearization technique assumes that the system model is known and all the states are available for feedback. But when the functions $f(z),\;g(z)$ in \eqref{flsys} are unknown, we cannot use feedback linearization to stabilize the system. Secondly, in most of the literature, the zero dynamics is assumed to have an asymptotically stable origin. We relax this assumption and investigate how the controllable subsystem can be controlled when the zero dynamics exhibits a periodic behavior and how the coupling between the internal and controllable subsystem affects the convergence of the controllable states. Thus, we propose a data-driven controller, which can stabilize the origin of the controllable subsystem, based on the following assumptions: 

 \begin{asum}
It is assumed that the normal form \eqref{plant1}-\eqref{plant4} is available, where the functions $\alpha(\cdot),\;\beta(\cdot),\;f_0(\cdot)$ are unknown. Also, we assume that $\beta$ is a constant and the unknown functions $\alpha(\cdot),\;f_0(\cdot)$  are Lipchitz in their arguments. \label{Assumption1} 
\end{asum}
 \begin{asum}
The zero-dynamics has a stable limit cycle, implying that $c_1\leq||\eta(t)||\leq c_2,\;\forall\;t\in\mathbb{R}^+$ and $c_1,\;c_2\in\mathbb{R}^+$. \label{Assumption2}
 \end{asum}
\section{Sampled-Data Equivalent of the controllable subsystem} \label{sdm}
 Consider the nonlinear SISO plant in the normal form \eqref{plant1}-\eqref{plant4}. Since the dynamics of this plant is assumed to be Lipschitz in their arguments as given in Assumption A\ref{Assumption1}, there exists a $T\in\mathbb{R}^{++}$ such that for all $t\in[t_0,\;t_0+T)$ the flow of each state of \eqref{plant2}-\eqref{plant4} denoted by $\phi_i(t)\triangleq\phi_i(\xi_0,\eta_0,u_0,t_0,t)$ using Taylor series expansion is given by:
 \begin{align}
     &\phi_i(t)=\sum_{j=i}^{j=\rho}({\xi_{j_0}\frac{t^{j-i}}{(j-i)!}})+\frac{t^{\rho-i+1}}{(\rho-i+1)!}(\alpha(\xi_0,\eta_0)+\beta u)\nonumber\\
     &+O_{(\xi,\eta,u-u_0)}(T^{\rho-i+2})
 \end{align}
 where $(\xi(t_0),\eta(t_0),u(t_0))=(\xi_0,\eta_0,u_0)\in\mathcal{D}$ are the initial conditions, where $\mathcal{D}$ is a compact set, $u_0=-\beta^{-1}\alpha(0)$ and $i\in\{1,2,\ldots\rho\}$. Henceforth, $O_{(\xi,\eta,u-u_0)}(\cdot)$ will be denoted as $O(\cdot)$, and will be mentioned explicitly when the arguments in the subscript changes.
 Setting $t_0=kT$, for a zero order hold assumption for the control input, that is, $u(kT+p)=u(kT)$, for all $p\in[kT,(k+1)T)$, we can derive the exact flow $\phi^e\triangleq col\{\phi_1^e(\cdot),\phi_2^e(\cdot),\ldots\}$ of the system  at time $(k+1)T$, given the initial condition at time $kT$, as:
  \begin{align}\label{discrete}
    &\xi_{i}(k+1)=\sum_{j=i}^{j=\rho}({\xi_j(k)\frac{T^{j-i}}{(j-i)!}})\nonumber\\
    &+\frac{T^{\rho-i+1}}{(\rho-i+1)!}(\alpha(\xi(k),\eta(k))+\beta u(k))+O(T^{\rho-i+2})
 \end{align}
 where the parameter $T$ is omitted. The expression given in \eqref{discrete} is valid at time $(k+1)T$, since the flow does not change over a zero-measure set. Neglecting the higher order terms, can approximate the flow $\phi^a\triangleq\phi^a(\xi(k),\eta(k),u(k),kT,(k+1)T)$:
   \begin{align}\label{approxdyn}
    &\xi_{i}(k+1)=\sum_{j=i}^{j=\rho}{\xi_j(k)\frac{T^{j-i}}{(j-i)!}}\nonumber\\
    &+\frac{T^{\rho-i+1}}{(\rho-i+1)!}(\alpha(\xi(k),\eta(k))+\beta u(k))
 \end{align}
 
\section{Feedback Linearization Based Controller} \label{cntr}
Now we design a nominal controller that guarantees asymptotic stability of the controllable subsystem via feedback linearization. Let $u(k)=\beta^{-1}(-\alpha(k)+v(\xi(k)),\;\; v(\xi(k))=K\xi$
where $K$ is a vector of feedback gains. Then the approximate dynamics \eqref{approxdyn} in closed loop will have the form $\xi(k+1)=A\xi(k)+Bv(k)$, where
\begin{align*}
    A=\begin{bmatrix} 
    1 & T &\hdots & \frac{T^{\rho-1}}{(\rho-1)!}\\ 
    0 & 1 &\hdots & \frac{T^{\rho-2}}{(\rho-2)!}\\
    \vdots&\vdots&\ddots&\vdots\\
    0 & 0 & \hdots & 1
    \end{bmatrix},\; B=\begin{bmatrix}\frac{T^{\rho}}{\rho}\\\frac{T^{(\rho-1)}}{(\rho-1)}\\\vdots\\T\end{bmatrix}
\end{align*}
The matrices $A$ and $B$ can be expanded as a series parameterized by $T$ as $A=I+A_1T+\hdots+A_{\rho-1}T^{\rho-1}$ and $B=B_1T+B_2T^2+\hdots+B_{\rho}T^{\rho}$, where the pair $(A_1,B_1)$ is controllable. Which implies, $\forall\;Q=Q^T\succ0$  $\exists$ $K$, $P=P^T\succ0$ such that $(A_1+B_1K)^TP+P(A_1+B_1K)=-Q$.
 Now computing the change in the Lyapunov function $ V(\xi(k))=\xi^T(k)P\xi(k)$ between each time step(omitting the argument $k$), we have:
\begin{align}
    V(\phi^a)-V(\xi)&=\xi^T(A+BK)^TP(A+BK)\xi-\xi^TP\xi\nonumber\\
     &=\xi^TT(A_1+B_1K)^TP\xi+\xi^TPT(A_1+B_1K)\xi\nonumber\\
     &-\xi^TP\xi+O_{{\xi}^2}(T^2)\nonumber\\
     &=-T\xi^TQ\xi+O_{{\xi}^2}(T^2)\nonumber\\
     &\leq-\lambda_{m}(Q)T{||\xi||}^2+O_{{\xi}^2}(T^2) \label{approxlyap}
\end{align}
From inequality \eqref{approxlyap}, we can say that if there exists a small enough sampling time $T$, then the change in Lyapunov function along the approximate flow of the system dynamics is negative definite. We will show that there exists such a small enough sampling time and derive the exact upper bound on the sampling time in Section \ref{main}. The derivation of inequality \eqref{approxlyap} is done similar to that of inequality $7.1$ in \cite{tabuada}.
Now, since $\alpha(\cdot)$ is unknown, we need an estimator that reconstructs $\alpha(\cdot)$ at each instant. Furthermore, since $\beta$ is assumed to be constant, estimating it once would suffice. We design a data-driven estimator in Section~\ref{est} for this purpose.
 \section{Data-Driven Estimator Design} \label{est}
\begin{figure}
      \centering
      \includegraphics[scale=0.5]{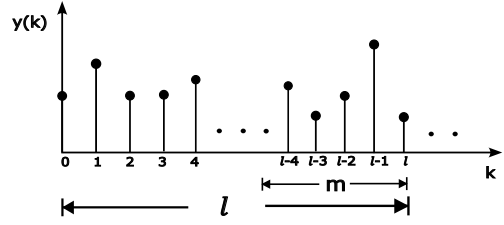}
      \caption{Initial data-collection period of length $l$ for estimating $\beta$ and data of length $m$ to estimate the value of $\alpha(k)$ at each instant beyond time $t=lT$.}
      \label{figure1}
   \end{figure}

Based on an extended plant model, we propose a data-driven estimation law which reconstructs $\alpha(\cdot)$ at each instant, and estimates the value of $\beta$. We set $\alpha(k)=\xi_{\rho+1}(k)$ to be an extended state that is assumed to remain constant between sampling instants for a small enough sampling time $T$. Then, the extended discretized plant dynamics is:
   \begin{align*}
       \xi_{i}(k+1)&=\sum_{j=i}^{j=\rho}({\xi_j(k)\frac{T^{j-i}}{(j-i)!}})\nonumber\\
     &+\frac{T^{\rho-i+1}}{(\rho-i+1)!}(\xi_{\rho+1}(k)+\beta u(k))+O(T^{\rho-i+2})\\
     \xi_{\rho+1}(k+1)&=\xi_\rho(k+1)+O(T)
 \end{align*}
for $i\in\{1,2\ldots\rho\}$. In the matrix representation, the above equations will have the form $\bar{\xi}(k+1)=\bar{A}\bar{\xi}(k)+\bar{B}\beta u(k)+w(k),\;y(k)=\bar{C}\bar{\xi}(k)$
where $w=col\{ O(T^{\rho+1}),\;O(T^{\rho}),\;\ldots\;O(T)\}$
    \begin{align*}
    \bar{A}&=\begin{bmatrix}
    1 & T & \frac{T^2}{2}& \hdots & \frac{T^{\rho}}{\rho!}\\
    0 & 1 & T& \hdots & \frac{T^{\rho-1}}{(\rho-)!}\\
    \vdots& \vdots & \ddots& \ddots & \vdots \\
    0 & 0 & 0 &\hdots & 1
    \end{bmatrix},\;\bar{B}=\begin{bmatrix}\frac{T^{\rho}}{\rho}\\\frac{T^{(\rho-1)}}{(\rho-1)}\\\vdots\\T\\0\end{bmatrix}
    \end{align*}
where $\bar{\xi}(k)$ is the extended state vector. An approximate model of this system, with the higher order term $w(k)$ neglected can be obtained as:
\begin{align}
     \bar{\xi}(k+1)&=\bar{A}\bar{\xi}(k)+\bar{B}\beta u(k), \;y(k)=\bar{C}\bar{\xi}(k)\label{approxmodelabcd1}
\end{align}
Essentially, this is what the observer ``thinks" the system looks like, and estimates the states based on this approximate model. Let $\{u(0),\;u(1)\ldots\;u(l-1)\}$ be a persistently exciting input sequence of order $2\rho+2$ such that $\text{rank}(\mathcal{Z}_0)=2\rho+2$, with $l\geq2\rho+2$, where 
\begin{align}
    \mathcal{Z}_0=\begin{bmatrix}H^l_{\rho+1}(y(-\rho-1))\\\hline H^l_{\rho+1}(u(-\rho-1))\end{bmatrix}+\begin{bmatrix}H^l_{\rho+1}(\bar{C}w(-\rho-1))\\\hline\vspace{0.02cm} \mathbb{O}\end{bmatrix}\label{z0}
\end{align}
and $H^l_{\rho+1}(u(-\rho-1)),\;H^l_{\rho+1}(y(-\rho-1)),\;H^l_{\rho+1}(\bar{C}w(-\rho-1))$ are the Hankel matrices of the input, output and the higher order terms respectively.
Let $\mathcal{Z}_1$ be the matrix $\mathcal{Z}_0$ with its entries shifted by one time-step.
\begin{prop}
    For the system \eqref{plant1}-\eqref{plant4}, with the Assumptions \eqref{Assumption1},\eqref{Assumption2}, the extended states can be estimated as:
\begin{align}
    \hat{\bar{\xi}}(k)={(\mathcal{O}^T\mathcal{O})}^{-1}\mathcal{O}(Y(k)+\hat{\beta} MU(k-1)) \label{ddest}
\end{align}
with an estimation error of $
    e_{\xi}:=\bar{\xi}-\hat{\bar{\xi}}=O(T)$, 
where $\hat{\beta}=(\mathcal{A}_{(\rho+1,\;2\rho+2)})\frac{\rho!}{T^{\rho}}$ and $\mathcal{A}_{(\rho+1,2\rho+2)}$ is the $(\rho+1,2\rho+2)^{th}$ element of the matrix $\mathcal{A}=\mathcal{Z}_1\mathcal{Z}_0^{\dagger}$. $Y(k)=col\{y(k),\;y(k-1), \ldots, y(k-m+1)\}$, $ U(k-1)=col\{0,\;u(k-1),\ldots,u(k-m+1)\}$,  where $m\geq\rho+1$, $\mathcal{O}={\begin{bmatrix}C^T&{(CA^{-1})}^T&\hdots&{(CA^{-m+1})}^T\end{bmatrix}}^T$ and
\begin{align*}
    M&=\begin{bmatrix}0 & 0 & 0 & \hdots& 0 \\
    0 & CA^{-1}B & 0 & \hdots &0\\
    0 & CA^{-2}B & CA^{-1}B &\hdots& 0\\
    \vdots& \ddots &\ddots &\vdots \\
    0 & CA^{-m+1}B & CA^{-m+2}B & \hdots& CA^{-1}B
    \end{bmatrix}_{m\times m} \label{MMatrix}
\end{align*}
\end{prop}
\begin{proof}
    From the approximate model given by \eqref{approxmodelabcd1}, we can observe that it is a discrete-time LTI system, with just one parameter unknown, i.e. $\beta$. Since the input sequence is persistently exciting,  we can use the data-based open-loop representation of a discrete-time LTI system given in Theorem 7 in \cite{formulabased}. Hence, the proof of the given expression for $\hat{\beta}$ will follow along the same lines as given in \cite{formulabased}. Now we have an estimate of $\beta$ obtained from the approximate discretized model. Next we show that the error in the estimate, i.e. $e_{\beta}=\beta-\hat{\beta}=O(T)$. Since $\hat{\beta}$ is obtained from an element in $\mathcal{A}$, 
\begin{align*}
    e_{\beta}\nonumber&=\bigg{|}\bigg{|}\mathcal{A}-\begin{bmatrix}H^l_{\rho+1}(y(-\rho))\\\hline H^l_{\rho+1}(u(-\rho))\end{bmatrix}{\begin{bmatrix}H^l_{\rho+1}(y(-\rho-1))\\\hline H^l_{\rho+1}(u(-\rho-1))\end{bmatrix}}^{\dagger}\bigg{|}\bigg{|}_F\nonumber\\
    &=O(T)
\end{align*}
where the last equality can be obtained by substituting $\mathcal{A}=\mathcal{Z}_1\mathcal{Z}_0^{\dagger}$, $\mathcal{Z}_0,\;\mathcal{Z}_1$ and the fact that $w(k)=O(T)$. Once $\hat{\beta}$ is obtained, we can write $Y(k)=\mathcal{O}\hat{\bar{\xi}}\nonumber\nonumber-\hat{\beta}MU(k-1)$, where $M$ is defined in \eqref{MMatrix}. Since the $\mathcal{O}$ matrix has linearly independent columns, due to the fact that $(A^{-1},C)$ is an observable pair, we rearrange the terms and take the pseudo-inverse of the matrix $\mathcal{O}$ to arrive at the estimator given in \eqref{ddest}. The error in this estimate $e_{\xi}=\bar{\xi}-\hat{\bar{\xi}}$ is due to the neglected $O(T)$ terms:
\begin{align}
     e_{\xi}&={(\mathcal{O}^T\mathcal{O})}^{-1}\mathcal{O}(Y(k)+\beta MU(k-1)+\mathcal{O}W(k-1))\nonumber \\
     &-{(\mathcal{O}^T\mathcal{O})}^{-1}\mathcal{O}(Y(k)+\hat{\beta} MU(k-1))=O(T)
\end{align}
where the last inequality is obtained as a result of $W(k-1)=O(T)$. This concludes the proof.\\
\end{proof} 
\begin{rem}
    We highlight the fact that the data collected during the initial $l$ time steps alone, must be persistently exciting to estimate $\beta$. Whereas the data window of length $m$, used to reconstruct the states and $\alpha(k)$ at each time step, need not be persistently exciting for $k\geq l+1$. A diagram showing the data-sampling intervals is given in Fig~\ref{figure1}.
\end{rem}
\section{Main Results} \label{main}
In this section, we present our main result, which guarantees the boundedness of the states, and in some special cases, the asymptotic stability of the controllable subsystem in the presence of periodic orbits in the zero-dynamics. 
\begin{theorem}\label{thm1}
Consider the plant given by \eqref{plant1}-\eqref{plant4}, and let assumptions \eqref{Assumption1},\eqref{Assumption2} hold. Let $\mathcal{D}$ be a compact set of initial conditions, that contains the origin. Then there exists a $T^*\in\mathbb{R}^+$ such that for all $T\in(0,T^*]$, the data-driven estimator \eqref{ddest} and a discrete control input, with a zero-order hold, given by $u(k)={\hat{\beta}^{-1}}(-\hat{\xi}_{\rho+1}(k)+v(\hat{\xi}(k)))$, where $v$ is a linear control law, guarantees $||\xi(k)||\leq b_1, \; \forall k\in\mathbb{N},\;\;||\xi(t)||\leq b_2,\; \forall t\geq0$
where $b_1,b_2\in\mathbb{R}^{++}$ depend on $\mathcal{D}$ , $T$ and $c_1,\;c_2$. 
Furthermore, if $\alpha(\cdot)$ in \eqref{plant3}, is a function of $\xi$ alone, then $\lim_{t\rightarrow \infty} \xi(t)=0$.
\end{theorem}
\begin{proof}
The estimator needs to wait for $l$ samples to arrive to estimate $\beta$, we first show that there exists a small enough sampling time such that the trajectories starting from any initial conditions that belong to a compact set remain in a compact set during this time interval. This time window will be termed the initial data-collection period.\\
\textbf{Initial Data-Collection Period}: 
The data-driven estimator needs a set of $l$ initial persistently exciting input data points to estimate the parameter $\beta$ and $m-1$, $m$ input and output data points respectively, which need not be persistently exciting to reconstruct the states, where $m\leq l$. The functions $\alpha(\cdot),\;f_0(\cdot)$ in $\eqref{plant2}-\eqref{plant4}$ are assumed to be Lipschitz in their arguments. Hence, we can guarantee that there exists a $T_1$ such that for all $T\in(0,T_1]$, with the initial conditions in $\mathcal{D}$, and the sequence of inputs $\{u(0),\;..u(l-1)\}$ belonging to a compact set $\mathcal{U}\subset\mathbb{R}$, each held constant between $t\in[kT,(k+1)T),\; k\in\{0,1,..l-1\}$,  the states and hence the output at each instant, will remain in a compact set $\mathcal{E}$ for all $k\in\{0,1,...l\}$. Now the data-driven estimator starts reconstructing the states from the time instant $t=lT$. 
\textbf{Evolution of states after time} $lT$: Let $\mathcal{F}$ be the smallest sub-level set of the Lyapunov function $V(\xi(t))$, containing the set $\mathcal{E}$. We now show that $\mathcal{F}$ is invariant.
Consider the evolution of the Lyapunov function $V$ along the exact flow of the system $\phi^e(\xi,\eta,u)$. Substituting $\phi^e$ from \eqref{discrete} and expanding, the difference in the Lyapunov function between two consecutive time steps along the exact flow satisfies:
\begin{align}\label{exact_approx}
    V(\phi^e)-V(\xi(k))\leq V(\phi^a)-V(\xi(k))+O(T^2)
\end{align}
 Now, since $\alpha$ is Lipschitz on the domain of interest, we can show that $O_{(\xi,\eta,u-u_0)}(T)=O_{\xi}(T)+O_{\eta}(T)$. Thus,
\begin{align}
     &V(\phi^e)-V(\xi(k))\leq V(\phi^a)-V(\xi(k))+O_{\xi^2}(T^2)+O_{\eta^2}(T^2)\nonumber\\
     &\leq-\lambda_{m}(Q)T{||\xi(k)||}^2+O_{\xi^2}(T^2)+O_{\eta^2}(T^2)\label{exactlyap}
\end{align}
The last inequality is obtained directly from \eqref{approxlyap}. Since $\hat{\xi}=\xi+O(T^2)$, \;$\hat{\xi}_{\rho+1}=\xi_{\rho+1}+O(T)$, \;$\hat{\beta}=\beta+O(T)$ and the fact that the Lyapunov function is quadratic, the inequality \eqref{exactlyap} remains unchanged. Hence,
\begin{align}
     &V(\phi^e)-V(\xi(k))\leq-\lambda_{m}(Q)T{||\xi(k)||}^2+O_{\xi^2}(T^2)+O_{\eta^2}(T^2)\nonumber\\
     &\leq -\lambda_{m}(Q)T{||\xi(k)||}^2+N_1T^2{||\xi(k)||}^2+N_2T^2{||\eta(k)||}^2\nonumber\\
     &\leq-\lambda_{m}(Q)T{||\xi(k)||}^2+N_2T^2{||\eta(k)||}^2\label{exactlyap2}
\end{align}
The inequality \eqref{exactlyap2} is obtained by choosing $T_2<\lambda_{m}(Q)/N_1$, for all $T\in(0,T_2]$.   
Since $||\eta(k)||$ is bounded (from Assumption \ref{Assumption2}), and for all $||\xi(k)||>\sqrt{(N_2T)/(\lambda_{m}(Q))}||\eta(k)||$, \eqref{exactlyap2} is negative definite. Thus proving the $\eta$-to-$\xi$ stability of the $\xi$ subsystem. (Here the $||\eta||$ term acts as an input to the $\xi$ dynamics). This also shows $\mathcal{F}$ is invariant for all $T\in(0,T_3]$, where $T_3=\min\{T_1,T_2\}$. Since $\mathcal{F}$ is compact, there exists a $b_1>0$ such that $||\xi(k)||\leq b_1,\;\forall k\in\mathbb{N}$. Invoking Theorem 5,  \cite{KLstability}, which ensures the uniform local input-to-state stability of the sampled data system, provided the corresponding discrete time system is uniformly locally input-to-state stable and uniformly bounded over $T$, we can guarantee that there exists a $b_2>0$ such that, $||\xi(t)||\leq b_2,\;\forall t\geq0$. Furthermore, if $\alpha$ is a function of $\xi(k)$ alone, the higher order terms are functions of $||\xi(k)||$ (i.e. $O_{\eta^2}(T^2)=0$) and \eqref{exactlyap} reduces to:
\begin{align}
     V(\phi^e)-V(\xi(k))&\leq -\lambda_{m}(Q)T{||\xi(k)||}^2
\end{align}
This guarantees asymptotic stability in discrete time. Now invoking Theorem 1 from \cite{KLstability}, which guarantees the uniform local asymptotic stability of the sampled data system, if the corresponding discrete time system is uniformly locally asymptotically stable and uniformly bounded over $T$, $\lim_{t\rightarrow0}\xi(t)=0$.
This concludes the proof.   
\end{proof}
\section{Discussion}
\textbf{Effect of Sampling Time and Noise:}  We can infer from \eqref{exactlyap2} that as $T$ decreases, the steady-state bound on the states also decreases. Thus the steady state error specifications in the controllable states can be met by choosing an appropriate sampling time. The presence of measurement noise alters the error in the state estimates as well as the estimate of $\beta$ to $e_{\xi}=e_{\beta}=O(T)+O_{\bar{\theta}}(T^{-\rho})$. Where, $\bar{\theta}=\sup_{t\in\mathbb{R}^+}||\theta(t)||$, and $\theta(t)$ is the measurement noise that appears additively in \eqref{plant4}. In this case, it can be shown that the states converge to a bounded set, with the bounds depending on the measurement noise, the sampling time, and the bounds on the zero dynamics.

\textbf{Minimum-Phase Systems}:      
    When the system is strictly minimum phase, $||\eta(k)||$ decreases to 0 and hence \eqref{exactlyap2} guarantees the asymptotic stability of the origin of \eqref{plant1}-\eqref{plant4}. Thus in the case of minimum phase systems, the results coincide with that of Theorem 8.1 in \cite{tabuada}.
    
\textbf{Robustness against perturbations in $\beta$}: The controller is seen to be fairly robust against small perturbations in $\beta$. If the function $\beta(\xi,\eta)=\beta_0+\Delta\beta(\xi,\eta)$, where $\beta_0$ is the nominal value of $\beta$ being estimated by the data-driven estimator, it can be shown that there exists a $\gamma$ such that $||\Delta\beta(\xi,\eta)||<\gamma$, for which the controller stabilizes the origin of the controllable subsystem. We shall incorporate this in the simulations to demonstrate the robustness of the controller. 
\section{Simulations}
\begin{figure}
      \centering
      \includegraphics[scale=0.3]{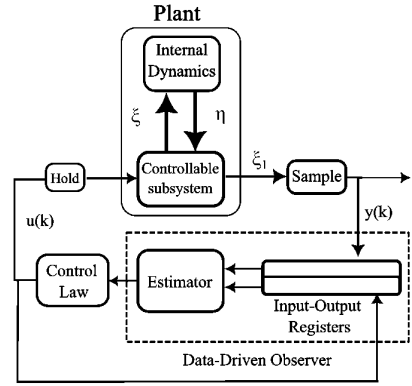}
      \caption{Block diagram representing various components of the closed-loop system.}
      \label{figure2}
   \end{figure}

\begin{figure}
      \centering
      \includegraphics[scale=0.45]{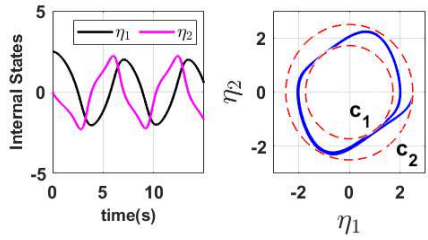}
      \caption{The internal state trajectories converge to a stable limit cycle.}
      \label{internal}
   \end{figure}

Consider the continuous time nonlinear system below:
\begin{align}
&\dot{\eta}_1=\eta_2,\;\dot{\eta}_2=-\eta_1+0.5(1-\eta_1^2)\eta_2+\xi_1\nonumber\\
    &\dot{\xi}_1=\xi_2,\;\;\dot{\xi}_2=-\mathrm{sin}(\xi_1)+2\xi_2+\delta\eta_1^2+(2+\Delta\beta(\xi,\eta))u\label{exm1_3}\nonumber\\ 
    &y=\xi_1\nonumber
\end{align}
where $\delta=1$ when there is a coupling between the subsystems and $\delta=0$ when there is no coupling between the subsystems. The limit cycle dynamics is assumed to be that of a Van der Pol oscillator. The relative degree of the system is $\rho=2$. The block diagram depicting the controller and the estimator implementation is given in Fig~\ref{figure2}.
Controller parameters were set to $T=0.02\;s$, $K=\begin{bmatrix}
-20 &-10
\end{bmatrix}$ to place the closed loop poles of the linearized controllable subsystem at $\lambda_1=-2.76,\;\lambda_2= -7.23$. The initial conditions for the states were $(1,0,2.5,0)$. We chose $l=8$ and $m=3$ for the data-driven estimator. The initial set of inputs were drawn from a normal distribution obtained using the ``randn()'' function in MATLAB. For the given initial conditions, the bounds on the internal states were found to be $c_1=1.72$ and $c_2=2.5$. The trajectories and the phase portrait of the internal states are given in Fig~\ref{internal}. 
       \begin{figure}
      \centering
      \includegraphics[scale=0.6]{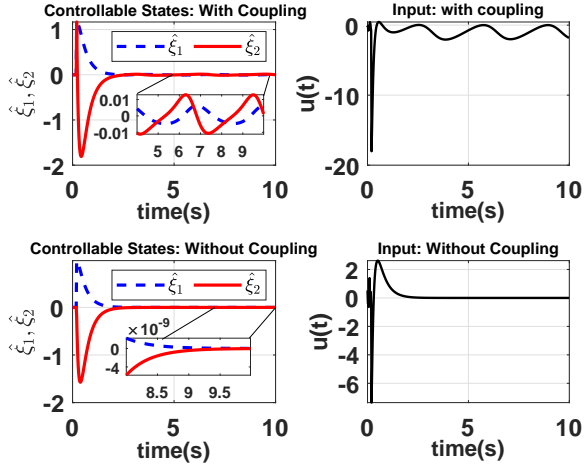}
      \caption{Effect of a coupling term between the subsystems on the stability of the controllable states.}
      \label{states}
   \end{figure}
     \begin{figure}
      \centering
      \includegraphics[scale=0.4]{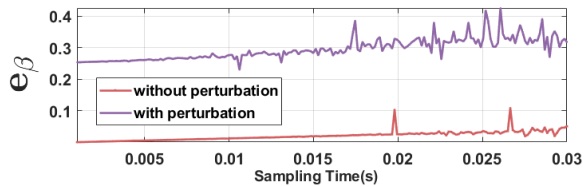}
      \caption{The red curve shows $e_{\beta}=O(T)$ when $\Delta\beta(\xi,\eta)=0$ and the purple plot shows how $e_{\beta}$ converges to a lower bound as the sampling time decreases due to the presence of perturbations in $\beta$.}
      \label{betaerrorplots}
   \end{figure}
   
\textbf{Case 1: When $\alpha(\cdot)$ is a function of both $\xi$ and $\eta$:} Let $\Delta\beta(\xi,\eta)=0.3\mathrm{sin}\xi_1$ be the perturbation in $\beta$. $\alpha(\xi,\eta)=-\mathrm{sin}\xi_1+2\xi_2+\eta_1^2$. The estimated value of $\beta$ is $\hat{\beta}=2.302$. The error in beta is large due to the presence of perturbations. Fig~\ref{states} shows that the controller stabilizes the origin of the controllable subsystem such that its trajectories converge to a small neighborhood around the origin.

\textbf{Case 2: When $\alpha(\cdot)$ is a function of $\xi$ alone:} 
In this case $\delta=0$, hence $\alpha(\xi,\eta)=-\mathrm{sin}\xi_1+2\xi_2$. It can be observed in Fig~\ref{states}, that the controller ensures asymptotic stability of the controllable states.

\textbf{Estimation error in $\beta$:} Validating the results derived, the error in estimating $\beta$, when $\Delta\beta(\xi,\eta)=0$ can be seen to be of $O(T)$. When $\Delta\beta(\xi,\eta)=0.3\mathrm{sin}\xi_1$, the error is seen to converge to a lower bound of $0.25$ as sampling time decreases. This is because the error due to perturbations in $\beta$ can't be eliminated by decreasing the sampling time. The plots are shown in Fig~\ref{betaerrorplots}. 
\section{CONCLUSIONS}
This article expanded the class of nonlinear systems for which data-driven techniques have been proposed, by investigating for the first time the scenario where the nonlinear systems have periodic orbits in the zero-dynamics. The article introduces a two-part data-driven estimator, which reconstructs the nonlinear terms for feedback linearization under the assumption that the initial set of data-points is persistently exciting. This data-driven estimator along with the feedback linearizing controller renders the closed-loop system disturbance-to-state stable. We also show that the estimation error scales linearly with the sampling time. In the future we look to extend the approach to a broader class of nonlinear systems and implement the data-driven controller for the head and torso stabilization of a biped robot.







\bibliographystyle{IEEEtran}
\bibliography{ref}

\begin{thebibliography}{10}
\providecommand{\url}[1]{#1}
\csname url@samestyle\endcsname
\providecommand{\newblock}{\relax}
\providecommand{\bibinfo}[2]{#2}
\providecommand{\BIBentrySTDinterwordspacing}{\spaceskip=0pt\relax}
\providecommand{\BIBentryALTinterwordstretchfactor}{4}
\providecommand{\BIBentryALTinterwordspacing}{\spaceskip=\fontdimen2\font plus
\BIBentryALTinterwordstretchfactor\fontdimen3\font minus
  \fontdimen4\font\relax}
\providecommand{\BIBforeignlanguage}[2]{{%
\expandafter\ifx\csname l@#1\endcsname\relax
\typeout{** WARNING: IEEEtran.bst: No hyphenation pattern has been}%
\typeout{** loaded for the language `#1'. Using the pattern for}%
\typeout{** the default language instead.}%
\else
\language=\csname l@#1\endcsname
\fi
#2}}
\providecommand{\BIBdecl}{\relax}
\BIBdecl

\bibitem{brunton}
U.~Fasel, E.~Kaiser, J.~N. Kutz, B.~W. Brunton, and S.~L. Brunton, ``Sindy with
  control: A tutorial,'' in \emph{2021 60th IEEE Conference on Decision and
  Control (CDC)}, 2021, pp. 16--21.

\bibitem{Dec}
J.~Lu, Z.~Yan, J.~Han, and G.~Zhang, ``Data-driven decision-making (d3m):
  Framework, methodology, and directions,'' \emph{IEEE Transactions on Emerging
  Topics in Computational Intelligence}, vol.~3, no.~4, pp. 286--296, 2019.

\bibitem{fault}
S.~Yin, H.~Gao, J.~Qiu, and O.~Kaynak, ``Fault detection for nonlinear process
  with deterministic disturbances: A just-in-time learning based data driven
  method,'' \emph{IEEE Transactions on Cybernetics}, vol.~47, no.~11, pp.
  3649--3657, 2017.

\bibitem{formulabased}
C.~De~Persis and P.~Tesi, ``Formulas for data-driven control: Stabilization,
  optimality, and robustness,'' \emph{IEEE Transactions on Automatic Control},
  vol.~65, no.~3, pp. 909--924, 2020.

\bibitem{ETC}
V.~Digge and R.~Pasumarthy, ``Data-driven event triggered control,'' \emph{20th
  European Control Conference}, 2022.

\bibitem{LQR}
G.~R. Gonçalves~da Silva, A.~S. Bazanella, C.~Lorenzini, and L.~Campestrini,
  ``Data-driven lqr control design,'' \emph{IEEE Control Systems Letters},
  vol.~3, no.~1, pp. 180--185, 2019.

\bibitem{minen}
G.~Baggio, V.~Katewa, and F.~Pasqualetti, ``Data-driven minimum-energy controls
  for linear systems,'' \emph{IEEE Control Systems Letters}, vol.~3, no.~3, pp.
  589--594, 2019.

\bibitem{dissipativity}
A.~Koch, J.~Berberich, and F.~Allgower, ``Provably robust verification of
  dissipativity properties from data,'' \emph{IEEE Transactions on Automatic
  Control}, pp. 1--1, 2021.

\bibitem{delay}
J.~G. Rueda-Escobedo, E.~Fridman, and J.~Schiffer, ``Data-driven control for
  linear discrete-time delay systems,'' \emph{IEEE Transactions on Automatic
  Control}, vol.~67, no.~7, pp. 3321--3336, 2022.

\bibitem{NLdatadriven}
\BIBentryALTinterwordspacing
R.~Strasser, J.~Berberich, and F.~Allgower, ``Data-driven control of nonlinear
  systems: Beyond polynomial dynamics,'' in \emph{2021 60th {IEEE} Conference
  on Decision and Control ({CDC})}.\hskip 1em plus 0.5em minus 0.4em\relax
  IEEE, dec 2021. [Online]. Available:
  \url{https://doi.org/10.11092Fcdc45484.2021.9683211}
\BIBentrySTDinterwordspacing

\bibitem{tabuada}
L.~Fraile, M.~Marchi, and P.~Tabuada, ``Data-driven stabilization of siso
  feedback linearizable systems,'' \emph{arXiv:2003.14240}, 2020.

\bibitem{walking}
P.-B. Wieber, R.~Tedrake, and S.~Kuindersma, \emph{Modeling and Control of
  Legged Robots}.\hskip 1em plus 0.5em minus 0.4em\relax Cham: Springer
  International Publishing, 2016.

\bibitem{Cluster}
T.~Menara, G.~Baggio, D.~S. Bassett, and F.~Pasqualetti, ``Stability conditions
  for cluster synchronization in networks of heterogeneous kuramoto
  oscillators,'' \emph{IEEE Transactions on Control of Network Systems},
  vol.~7, no.~1, pp. 302--314, 2020.

\bibitem{slotine}
J.-J.~E. Slotine, W.~Li \emph{et~al.}, \emph{Applied nonlinear control}.\hskip
  1em plus 0.5em minus 0.4em\relax Prentice hall Englewood Cliffs, NJ, 1991,
  vol. 199, no.~1.

\bibitem{isidori}
A.~Isidori, \emph{Nonlinear control systems: an introduction}.\hskip 1em plus
  0.5em minus 0.4em\relax Springer, 1985.

\bibitem{KLstability}
D.~Nesic, A.~Teel, and E.~Sontag, ``Formulas relating kl stability estimates of
  discrete-time and sampled-data nonlinear systems,'' \emph{Systems \& Control
  Letters}, vol.~38, no.~1, pp. 49--60, 1999.

\end{thebibliography}

\end{document}